\documentclass{iopart}
\usepackage{graphicx}

\begin{document}
\title[ ]{Magnetic field dependence of the antiferromagnetic phase transitions in Co-doped YbRh$_2$Si$_2$}

\author{Christoph Klingner, Cornelius Krellner, and Christoph Geibel}
\ead{krellner@cpfs.mpg.de}
\address{Max Planck Institute for Chemical Physics of Solids, D-01187 Dresden, Germany}

\begin{abstract}
We present first specific-heat data of the alloy Yb(Rh$_{1-x}$Co$_x$)$_2$Si$_2$ at intermediate Co-contents $x=0.18$, 0.27, and 0.68. The results already point to a complex magnetic phase diagram as a function of composition. Co-doping of YbRh$_2$Si$_2$ ($T_N^{x=0}=72$\,mK) stabilizes the magnetic phase due to the volume decrease of the crystallographic unit cell. The magnetic phase transitions are clearly visible as pronounced anomalies in $C^{4f}(T)/T$ and can be suppressed by applying a magnetic field. Going from $x=0.18$ to $x=0.27$ we observe a change from two mean-field (MF) like magnetic transitions at $T_N^{0.18}=1.1$\,K and $T_L^{0.18}=0.65$\,K to one sharp $\lambda$-type transition at $T_N^{0.27}=1.3$\,K. Preliminary measurements under magnetic field do not confirm the field-induced first-order transition suggested in the literature. For $x=0.68$ we find two transitions at $T_N^{0.68}=1.14$\,K and $T_L^{0.68}=1.06$\,K. 
\end{abstract}

\section{Introduction}
The heavy-fermion system YbRh$_2$Si$_2$ is a clean, stoichiometric, and well-characterized metal situated extremely close to an antiferromagnetic (AFM) quantum critical point (QCP). This leads to pronounced non-Fermi-liquid behavior in transport and thermodynamic properties, such as the divergence of the electronic Sommerfeld coefficient $\gamma=C^{4f}/T$, and a linear-in-T resistivity \cite{Trovarelli:2000a}. The observed temperature and field dependences of many quantities disagree with the expectation for the three-di\-men\-sio\-nal spin-density-wave scenario and point to the new, still developing concept of local quantum criticality, where the Kondo effect is critically destroyed at the QCP (for a review see Ref.~\cite{Gegenwart:2008}). The recent discoveries of an additional energy scale vanishing at the QCP which does neither correspond to the N\'eel temperature nor to the upper boundary of the Fermi-liquid region  \cite{Gegenwart:2007} and a large critical exponent $\alpha=0.38$ at the AFM phase transition observed in low-temperature specific-heat measurements on a single crystal of superior quality \cite{Krellner:2009} have once again boosted the interest in YbRh$_2$Si$_2$. The latter observation triggered strong theoretical effort to explain the violation of critical universality in terms of a (quantum) tricritical point \cite{Misawa:2009, Shaginyan:2009}. In this scenario, Misawa \textit{et al.}~\cite{Misawa:2009} proposed the existence of a tricritical point for YbRh$_2$Si$_2$ under pressure and magnetic field at finite temperatures. Experimentally, this part of the phase diagram is easiest to explore using chemical pressure as will be discussed below.

The magnetic ordering of YbRh$_2$Si$_2$ ($T_N = 72$\,mK) is stabilized by applying pressure as expected for Yb-Kondo lattice compounds \cite{Plessel:2003}. The complementary method of substituting smaller isoelectronic Co for Rh results in chemical pressure allowing a detailed investigation of the magnetic phase diagram and the physical behavior of the stabilized AFM ordered state. Therefore, a thorough understanding of the physical properties of Yb(Rh$_{1-x}$Co$_x$)$_2$Si$_2$ are of great interest in order to understand the phenomena at the QCP in YbRh$_2$Si$_2$. Very recently, it was shown that for $x=0.07$ the signature of the Kondo breakdown is located within the magnetically ordered phase leading to a detaching of the AFM QCP from the Fermi-surface reconstruction \cite{Friedemann:2009a}.

\section{Experimental}
Single crystals of the alloy series Yb(Rh$_{1-x}$Co$_x$)$_2$Si$_2$ were grown from In flux, analogous to the stoichiometric samples of superior quality \cite{Krellner:2009}. However, the statistical substitution of Rh with Co leads to larger disorder compared to YbRh$_2$Si$_2$. A thorough investigation of the complete doping series, including x-ray diffraction measurements, magnetic susceptibility, electrical resistivity, and specific heat was performed and will be published separately \cite{Klingner:2009}. The Co-concentration was accurately determined using energy dispersive x-ray spectra of the polished single crystals. This real Co-content will be used for $x$ throughout the manuscript.

In this contribution, we present the magnetic field dependence of the specific heat in the vicinity of the AFM transition for three representative samples with $x=0.18$, 0.27, and 0.68, respectively. The specific heat was determined with a commercial (Quantum Design) physical property measurement system (PPMS) equipped with an $^3$He-insert, using a standard heat-pulse relaxation technique. The 4$f$ contribution to the specific heat, $C^{4f}$, was obtained by subtracting the non-magnetic one, $C_{\rm Lu}$, from the measured specific heat, $C_{\rm meas}$. $C_{\rm Lu}$ was determined by measuring the specific heat of the non-magnetic reference sample LuRh$_2$Si$_2$ below 10\,K \cite{Ferstl:2007}. Since $C_{\rm Lu}$ at 1\,K contributes only to 1\,\% of $C_{\rm meas}$, the synthesis and measurement of the appropriate reference systems Lu(Rh$_{1-x}$Co$_x$)$_2$Si$_2$ was not necessary. The nuclear quadrupolar contribution to the specific heat was not subtracted but is negligible in the investigated temperature and magnetic field range.

\begin{figure}[t]
\begin{center}
\includegraphics[width=10cm]{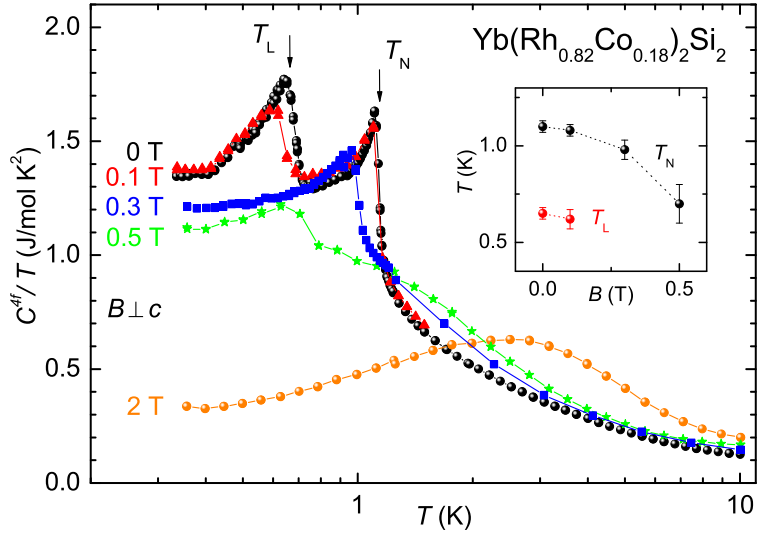}
\caption{$C^{4f}/T$ vs $T$ of a single crystal with $x=0.18$. For $B=0$\,T, the two magnetic transitions $T_N$ and $T_L$ are indicated by arrows. Both shift to lower temperature with increasing $B$, as indicated in the inset. The   transitions are suppressed at $B=2$\,T.} 
\label{fig1}
\end{center}
\end{figure}

\begin{figure}[t]
\begin{center}
\includegraphics[width=10cm]{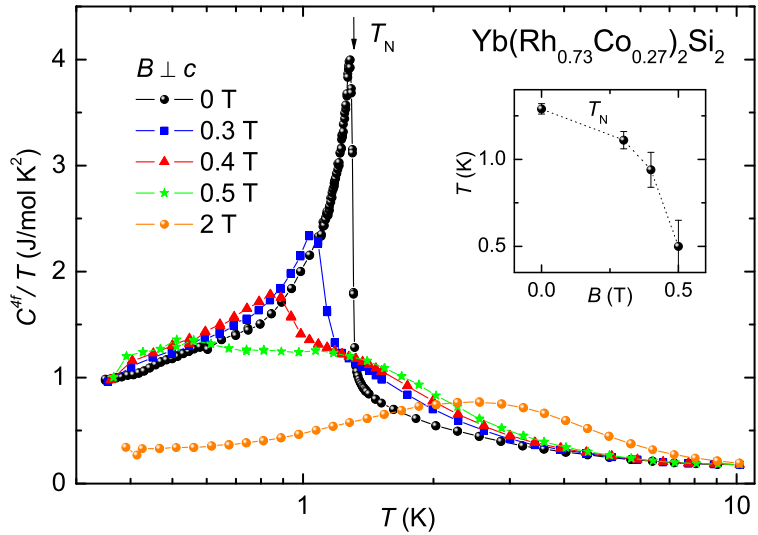}
\caption{$C^{4f}/T$ vs $T$ of a single crystal with $x=0.27$. For $B=0$\,T, only one magnetic transition is visible at $T_N$, indicated by an arrow. $T_N$ shifts to lower temperature with increasing $B$, as indicated in the inset. The transitions are suppressed at $B=2$\,T.} 
\label{fig2}
\end{center}
\end{figure}

\begin{figure}[t]
\begin{center}
\includegraphics[width=10cm]{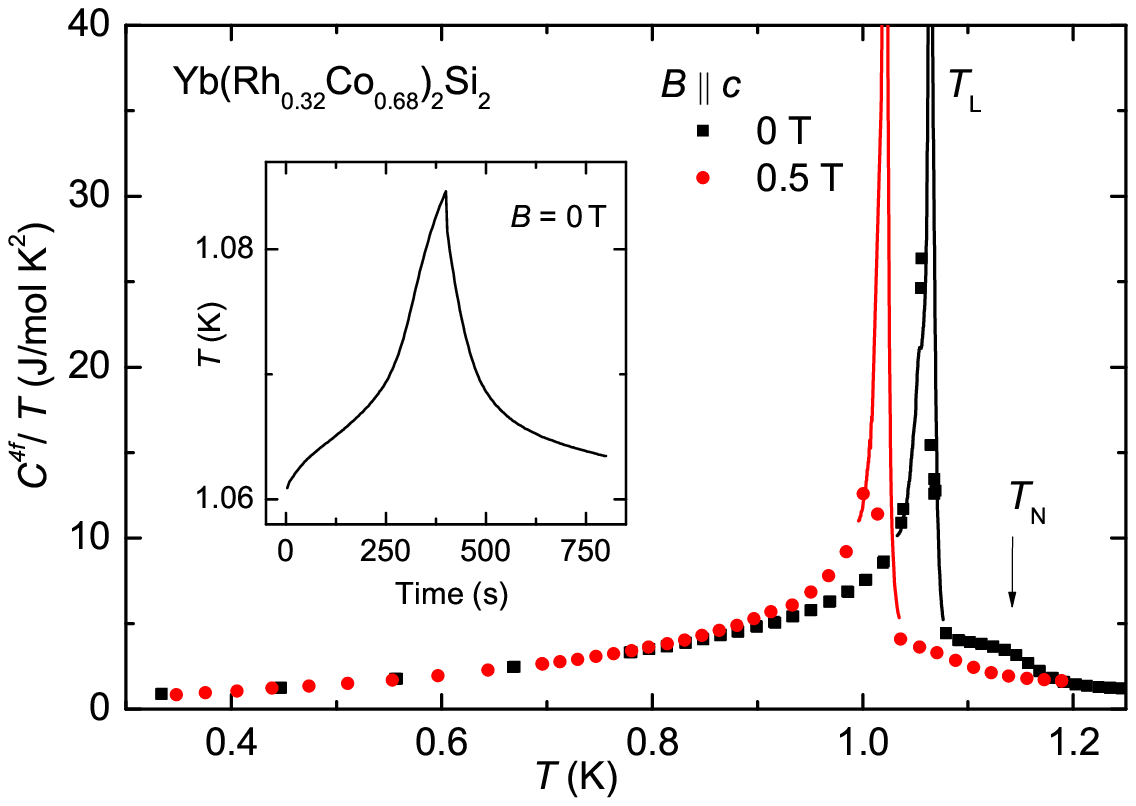}
 \caption{Specific heat of a single crystal with $x=0.68$ for $B=0$ and $B=0.5$\,T ($B\parallel c$). In zero field a broad anomaly is visible at $T_N^{0.68}=1.14$\,K, followed by a very sharp first-order transition at $T_L^{0.68}=1.06$\,K. Both transitions shift to lower temperatures in field. The solid data points depicted values obtained by using the standard $2\tau$-model to fit the relaxation-time curve, which cannot be used to extract values around a first-order phase transition. Instead, the specific heat around $T_L$ (solid lines) was directly determined from the heating curve of the relaxation-time measurement (shown in the inset) as described in Ref.~\cite{Lashley:2003}.}
\label{fig3}
\end{center}
\end{figure}

\section{Results}
In Fig.~\ref{fig1} the $4f$ part of the specific heat is shown for a single crystal with $x=0.18$. At zero magnetic field, two distinct anomalies are visible at $T_N^{0.18}=1.1$\,K and $T_L^{0.18}=0.65$\,K indicated by the arrows. We note, that these anomalies at $T_{N/L}$ are more mean-field (MF) like compared to the very sharp $\lambda$-peak at $T_N$ for $x=0$ \cite{Krellner:2009}. For $x=0.18$, both transitions shift to lower temperatures when a magnetic field is applied along the basal plane, which is the magnetic easy direction. This is presented in the inset of Fig.~\ref{fig1}, where a preliminary magnetic phase diagram is drawn from the specific-heat data. Magnetic measurements for $x=0.07$ have confirmed that these subsequent phase transitions, $T_N$ and $T_L$, are of magnetic origin, as they exhibit pronounced, AFM-like anomalies in the magnetic susceptibility \cite{Friedemann:2009a}. Below $T_L^{0.18}$, $C^{4f}/T$ becomes constant presenting a strongly enhanced Sommerfeld coefficient, $\gamma_{\rm 350\,mK}^{0.18}\approx 1.35$\,J/mol\,K$^2$, most probably due to pronounced Kondo interactions \cite{Desgranges:1982}.  At $B=2$\,T, both magnetic transitions are completely suppressed and the Sommerfeld coefficient amounts to only $\gamma\approx 0.3$\,J/mol\,K$^2$. This field curve presents a broad maximum around 2.5\,K which most probably is due to the Zeeman-splitting of the doublet ground state, and was also observed for $x=0$ \cite{Gegenwart:2006}. It is important to note that the MF-like transitions get even broader when a magnetic field is applied, in contrast to what is expected in the quantum-tricritical-point scenario proposed by Misawa \textit{et al.}~\cite{Misawa:2009}. The chemical pressure for $x=0.18$ corresponds to $p=3$\,GPa, well above the critical pressure $p_t$ in Ref.~\cite{Misawa:2009}. However, we could not observe any sign of a first-order transition above 350\,mK, which should be clearly visible in the specific heat, as e.g., observed for the lower magnetic transition $T_L^{0.68}$ for $x=0.68$ discussed below.

In Fig.~\ref{fig2}, we present the specific heat of a single crystal with $x=0.27$. In contrast to the $x=0.18$ sample (Fig.~\ref{fig1}), only one sharp anomaly is distinguishable at $T_N^{0.27}=1.3$\,K. However, the entropy reached at 10\,K is only slightly larger compared to the $x=0.18$ sample as will be analyzed separately \cite{Klingner:2009}, while the Sommerfeld-coefficient at 350\,mK is only slightly smaller, $\gamma_{\rm 350\,mK}^{0.27}\approx 1.0$\,J/mol\,K$^2$. The magnetic field dependence is analogous to what is observed for the $x=0.18$ crystal: $T_N^{0.27}$ decreases with increasing field (see inset of Fig.~\ref{fig2}) without any signature of a first-order phase transition and the data for $B=2$\,T shows again a broad maximum around 2.5\,K, with a reduced Sommerfeld coefficient, $\gamma\approx 0.3$\,J/mol\,K$^2$.

Fig.~\ref{fig3} shows the specific heat for a sample with higher Co-concentration, $x=0.68$. Here, two phase transitions are visible at $T_N^{0.68}=1.14$\,K and $T_L^{0.68}=1.06$\,K. The first one is reflected as a broad MF-like anomaly around $T_N^{0.68}$, whereas the second one peaks up at $T_L^{0.68}$ with clear signatures of a first-order phase transition. For tracing such a sharp peak using the software of the PPMS, instead of determining a single $C(T)$ value by fitting the whole relaxation curve with an exponential function (data points in Fig.~\ref{fig3}), it is more appropriate to calculate a continuous $C(T)$ curve from the time derivative of the relaxation curve directly (solid lines in Fig.~\ref{fig3}). The relaxation curve across the transition is shown in the inset of Fig.~\ref{fig3}. One immediately notices the shoulder in the heating part which is a direct evidence for a thermal arrest and thus for a first-order transition \cite{Lashley:2003}. Preliminary magnetic field dependent data are also shown in Fig.~\ref{fig3} (red curve). Both transitions are shifted to lower temperatures indicative of AFM phase transitions in agreement what was presented for lower Co-concentrations. However, it is difficult to compare quantitatively the magnetic field dependence with the data of the lower concentrations, as for $x=0.68$ the magnetic field was applied parallel to the $c$-axis which is the magnetic hard direction.

\section{Conclusion}
In summary, we have presented specific-heat measurements for three selected Co-concentrations of the series Yb(Rh$_{1-x}$Co$_x$)$_2$Si$_2$ with $x=0.18$, 0.27, and 0.68. The magnetic phase transitions are clearly visible as pronounced anomalies in the temperature dependence of the specific heat and can be suppressed by applying a magnetic field, without any signatures of a field-induced first-order transition. Going from $x=0.18$ to $x=0.27$ we observe a change from two MF-like magnetic transitions at $T_N^{0.18}=1.1$\,K and $T_L^{0.18}=0.65$\,K to one $\lambda$-type transition at $T_N^{0.27}=1.3$\,K. For $x=0.68$ we observe two transitions at $T_N^{0.68}=1.14$\,K and $T_L^{0.68}=1.06$\,K. The upper one is MF-like while the lower one presents the characteristics of a first-order phase transition. These results of varying magnetic phase transitions suggest a complex magnetic phase diagram for the series Yb(Rh$_{1-x}$Co$_x$)$_2$Si$_2$. In a forthcoming publication \cite{Klingner:2009} we will analyze and discuss the different types of order and compare them with results on YbRh$_2$Si$_2$ under hydrostatic pressure.

%\ack
The authors thank U. Burkhardt and P. Scheppan for energy dispersive x-ray analysis of the samples and R. Weise for
technical assistance in sample preparation. R. Borth, M.~Brando, S.~Friedemann, P.~Gegenwart, S.~Lausberg, M.~Nicklas, N.~Oeschler, L.~Pedrero, J.~Sichelschmidt, F.~Steglich, and O.~Stockert are acknowledged for valuable discussions.


\section{References}

\begin{thebibliography}{10}
\expandafter\ifx\csname url\endcsname\relax
  \def\url#1{{\tt #1}}\fi
\expandafter\ifx\csname urlprefix\endcsname\relax\def\urlprefix{URL }\fi
\providecommand{\eprint}[2][]{\url{#2}}
% Bibliography created with iopart-num v2.0
% /biblio/bibtex/contrib/iopart-num

\bibitem{Trovarelli:2000a}
Trovarelli O \textit{et al.} 2000 {\em Phys. Rev. Lett.\/} {\bf 85} 626

\bibitem{Gegenwart:2008}
Gegenwart P, Si Q and Steglich F 2008 {\em Nature Phys.\/} {\bf 4} 186

\bibitem{Gegenwart:2007}
Gegenwart P \textit{et al.} 2007 {\em Science\/} {\bf 315} 969

\bibitem{Krellner:2009}
Krellner C, Hartmann S, Pikul A, Oeschler N, Donath J~G, Geibel C, Steglich F
  and Wosnitza J 2009 {\em Phys. Rev. Lett.\/} {\bf 102} 196402

\bibitem{Misawa:2009}
Misawa T, Yamaji Y and Imada M 2009 {arXiv:0905.2046v1}

\bibitem{Shaginyan:2009}
Shaginyan V~R, Amusia M~Y and Popov K~G 2009 {arXiv:0905.1871v1}

\bibitem{Plessel:2003}
Plessel J, Abd-Elmeguid M, Sanchez J, Knebel G, Geibel C, Trovarelli O and
  Steglich F 2003 {\em Phys. Rev. B\/} {\bf 67} 180403

\bibitem{Friedemann:2009a}
Friedemann S \textit{et al.} 2009 {\em Nature Phys.\/}  {\bf 5} 465

\bibitem{Klingner:2009}
Klingner C \textit{et al.}, to be published

\bibitem{Ferstl:2007}
Ferstl J 2007 Pd.D. thesis, TU Dresden

\bibitem{Desgranges:1982}
Desgranges H U and Schotte K D 1982 {\em Phys. Lett.\/}
  {\bf 91A} 240

\bibitem{Gegenwart:2006}
Gegenwart P \textit{et al.} 2006 {\em New J. Phys.\/}
  {\bf 8} 171

\bibitem{Lashley:2003}
Lashley J~C \textit{et al.} 2003 {\em
  Cryogenics\/} {\bf 43} 369

\end{thebibliography}
\end{document}